\documentclass[12pt,english]{revtex4}
\usepackage[T1]{fontenc}
\usepackage[latin1]{inputenc}

\makeatletter

\usepackage{babel}
\makeatother
\begin{document}

\title{New physics upper bound on the branching ratio of 
$B_{s}\rightarrow l^{+}l^{-}$}

\author{Ashutosh Kumar Alok and S. Uma Sankar}

\affiliation{Department of Physics, Indian Institute of Technology,
Bombay, Mumbai-400076, India}

\today

\begin{abstract}
We consider the most general new physics effective Lagrangian 
for $b\rightarrow sl^+l^-$.
We derive the upper limit on the branching ratio for the processes 
$B_{s}\rightarrow l^+l^-$ where $l=e,\mu$, subject to the current 
experimental bounds on related processes, $B\rightarrow Kl^+l^-$ and
$B\rightarrow K^*l^+l^-$. If the new physics interactions are of 
vector/axial-vector form, the present measured rates for 
$B \rightarrow (K,K^*) l^+ l^-$ constrain $B(B_s\rightarrow l^+l^-)$
to be of the same order of magnitude as their respective Standard Model
predictions. On the other hand, if the new physics interactions are of
scalar/pseudo-scalar form, $B \rightarrow (K,K^*) l^+ l^-$ rates do
not impose any constraint on $B_s \rightarrow l^+ l^-$ and the branching 
ratios of these decays can be as large as present experimental upper bounds. 
If future experiments measure $B(B_s \rightarrow l^+ l^-)$ to be 
$\geq 10^{-8}$ then the new physics giving rise to these decays has to 
be of the scalar/pseudo-scalar form.
\end{abstract}

%\showpacs{13.20.He,12.15.Mm}

\maketitle

The rare decays of B mesons involving flavour changing neutral interaction
(FCNI) $b\rightarrow s$ has been a topic of great interest for long.
Not only will it subject the standard model (SM) to accurate tests
but will also put strong constraints on several models beyond the
SM. In the SM, FCNI occur only via one or more loops. Thus the rare
decays of B mesons will provide useful information about the higher-order
effects of the SM. Recently, the very high statistics experiments at 
B-factories have measured non-zero values for the branching
ratios for the FCNI processes $B\rightarrow(K,K^*)l^+l^-$ 
\cite{belle-03, babar-03}, 
\begin{eqnarray}
Br(B\rightarrow Kl^{+}l^{-}) & = & 
(4.8_{-0.9}^{+1.0}\pm0.3\pm0.1)\times10^{-7}, \nonumber \\
Br(B\rightarrow K^{*}l^{+}l^{-}) & = & 
(11.5_{-2.4}^{+2.6}\pm0.8\pm0.2)\times10^{-7}.
\end{eqnarray}
These branching ratios are close to the values predicted by the SM
\cite{deshpande-88}. However, the SM predictions for them contain 
about $\sim15\%$ uncertainty coming from the hadronic form factors.  
Still, it is worth considering what constraints these measurements 
impose on other related processes.

The effective Lagrangian for the four fermion process 
$b\rightarrow sl^+l^-$
gives rise to the exclusive semi-leptonic decays such as 
$B\rightarrow Kl^+l^-$
and $B\rightarrow K^*l^+l^-$ and also to purely leptonic decays
$B_s\rightarrow l^+l^-$, where $l=e,\mu$. (From here onwards,
the symbol $l$ represents either $e$ or $\mu$.) 
Relation between semi-leptonic and purely leptonic B-decays, 
arising from FCNI generated by heavy $Z'$ boson exchange, 
was briefly considered in \cite{langplum-00,barglang-04}.
The SM predictions for the branching ratios for the decays
$B_s\rightarrow e^+e^-$ and $B_s\rightarrow\mu^+\mu^-$
are $(7.58\pm3.5)\times10^{-14}$ and $(3.2\pm1.5)\times10^{-9}$
respectively \cite{buras-01}. The large uncertainy in the SM prediction
for these branching ratios arises due to the $12 \%$ uncertainty in
the $B_s$ decay constant and $10 \%$ uncertainty in the measurement of
$V_{ts}$. These branching ratios 
have been calculated in various new physics models. In models with 
$Z'$-mediated FCNI, one has $B(B_s\rightarrow\mu^+\mu^-)
<5.8\times10^{-8}$ \cite{london-97} which is about 20 times larger
than the SM prediction. 
Due to the increased precision in the measurement of
$B(B \rightarrow (K,K^*) l^+ l^-)$, this bound can be improved and
the present calculation attempts to do so. 
$B(B_s \rightarrow l^+ l^-)$ are also calculated in multi Higgs 
doublet models. These models are classified into two types. 
In the first type, there is natural flavour conservation
(NFC) and there are no FCNI at tree level. In such models, there is
an additional loop contribution to FCNI, where a charged Higgs boson
exchange replaces the SM W-exchange. In a two Higgs doublet
model with NFC, branching ratio for $B_s\rightarrow\mu^+\mu^-$ 
$\geq10^{-8}$ is possible \cite{hewett-89}. In the second type,
flavour changing processes do occur at tree level, mediated by 
flavour changing neutral scalars (FCNS's). 
In such models also a branching ratio of about
$10^{-8}$ for $B_s\rightarrow\mu^+\mu^-$ can be achieved 
\cite{london-97}. From the experimental side, at present, there exist 
only upper bounds 
$B(B_{s}\rightarrow e^{+}e^{-}) < 5.4\times10^{-5}$ \cite{l3-97} and 
$B(B_{s}\rightarrow\mu^{+}\mu^{-}) < 5.0\times10^{-7}$ \cite{abazov-04}.

In this paper, we consider the most general four fermion effective 
Lagrangian for $b\rightarrow sl^+l^-$ transition due to new physics. 
We derive upper bounds on the branching ratios for $B_s\rightarrow e^+e^-$
and $B_s\rightarrow\mu^+\mu^-$ by demanding that the predictions
of this new physics Lagrangian for $B\rightarrow K^{*}l^+l^-$
and $B\rightarrow Kl^+l^-$ should be consistent with the current
experimental values.

The most general effective Lagrangian for $b\rightarrow sl^+l^-$ 
transitions due to new physics can be written as,
\begin{equation}
L_{eff}\,(b\rightarrow sl^+l^-)=L_{VA} + L_{SP} + L_T
\end{equation}
where, $L_{VA}$ contains vector and axial-vector couplings,
$L_{SP}$ contains scalar and psuedo-scalar couplings and
$L_T$ contains tensor couplings. $L_T$ does not contribute
to $B_s \rightarrow l^+ l^-$ because $\langle 0| \bar{s}
\sigma^{\mu \nu} b | B_s(p_B) \rangle = 0$. Hence we
will drop it from further consideration.

First we will assume that the new physics Lagrangian contains 
only vector and axial-vector couplings. We parametrize it as
\begin{equation}
L_{VA}\,(b\rightarrow sl^+l^-)=\frac{G_{F}}{\sqrt{2}}
\left(\frac{\alpha}{4\pi s_W^2}\right)
\bar{s}(g_V+g_A\gamma_5)\gamma_\mu  \,
\bar{l}(g^{'}_V+g^{'}_A\gamma_5)\gamma^\mu l.\label{effL}
\end{equation}
Here the constants $g$ and $g'$ are the effective couplings which 
charecterise the new physics. 
From the above equation, we get $B_s \rightarrow l^+ l^-$ 
matrix element to be  
\begin{equation}
M\,(B_{s}\rightarrow l^{+}l^{-})=(g_{A}g_{A}^{'})\frac{G_{F}}{\sqrt{2}}
\left(\frac{\alpha}{4\pi s_{W}^{2}}\right)
\langle 0\left|\overline{s}\gamma_{5}\gamma_{\mu}b\right|B_s\rangle 
\langle l^+ l^-\left|\overline{l}\gamma_{5}\gamma^{\mu}l\right|0\rangle.
\label{bsllbva}
\end{equation}
Only the axial vector parts contribute for both the hadronic and
leptonic parts of the matrix element. Substituting 
$\langle 0\left|\overline{s}\gamma_{5}\gamma_{\mu}b\right|B_s\rangle 
\,=\,-if_{B_{s}}p_{B\mu},$
in Eq.~(\ref{bsllbva}) we get 
\begin{equation}
M\,(B_{s}\rightarrow l^{+}l^{-})=
-i2m_{l}f_{B_{s}}(g_{A}g_{A}^{'})
\frac{G_{F}}{\sqrt{2}}\left(\frac{\alpha}{4\pi s_{W}^{2}}\right)
\bar{u}(p_{l})\gamma_{5}v(p_{\bar{l}}).
\end{equation}
As we are considering only vector and axial
vector currents, helicity suppression is still operative for the 
$B_{s}\rightarrow l^{+}l^{-}$ decay amplitude.
The calculation of the decay rate gives 
\begin{equation}
\Gamma_{NP}(B_{s}\rightarrow l^{+}l^{-})\,=\,
\frac{G_{F}^{2}f_{B_{s}}^{2}}{8\pi}
\left(\frac{\alpha}{4\pi s_{W}^{2}}\right)^{2}
(g_{A}g_{A}^{'})^{2}m_{B_{s}}m_{l}^{2}. \label{npbs}
\end{equation}
Thus the decay rate depends upon the value of $(g_{A}g_{A}^{'})^{2}$.
To estimate the value of $(g_{A}g_{A}^{'})^{2}$, we consider the
related decays $B\rightarrow K^{*}l^{+}l^{-}$ and 
$B\rightarrow Kl^{+}l^{-}$,
which also receive contributions from the effective Lagrangian in
Eq.~(\ref{effL}). In deriving Eq.~(\ref{npbs}), we dropped 
terms proportional to $m_l^2/m_B^2$, as their contribution is 
negligible. We will make the same approximation in calculating 
the decay width of semi-leptonic modes also. 

We first consider the process $B\rightarrow K^{*}l^{+}l^{-}$.
Here we will have to calculate the following hadronic matrix elements 
\cite{deshpande-88}:
\begin{eqnarray}
\langle K^{*}(p_{K^{*}})\left|\overline{s}\gamma_{\mu}b\right|B(p_{B})\rangle 
& \,=\, & i\epsilon_{\mu\vartheta\lambda\sigma}\epsilon^{\nu}
(p_{K^{*}})(p_{B}+p_{K^{*}})^{\lambda}(p_{B}-p_{K^{*}})^{\sigma}V(q^{2}) 
\nonumber \\
\langle K^{*}(p_{K^{*}})\left|\overline{s}\gamma_{5}\gamma_{\mu}b\right|
B(p_{B})\rangle & \,=\, & 
\epsilon_{\mu}(p_{K^{*}})(m_{B}^{2}-m_{K^{*}}^{2})A_{1}(q^{2})-
(\epsilon.q) (p_B + p_{K^{*}})_\mu A_2 (q^2)
\end{eqnarray}
where $q=p_{l^{+}}+p_{l^{-}}$. In the above equation, a term proportional
to $q_\mu$ is dropped because its contribution to the decay rate is 
proportional to $m_l^2/m_B^2$.
It is assumed that the $q^{2}$ dependence of these form factors is
well described by a pole fit:
\begin{eqnarray}
V(q^{2}) & = & \frac{V}{(m_{B}+m_{K^{*}})(1-q^{2}/m_{B}^{2})} \nonumber \\
A_{i}(q^{2}) & =& \frac{A_{i}}{(m_{B}+m_{K^{*}})(1-q^{2}/m_{B}^{2})}.
\nonumber 
\end{eqnarray}
The decay rate is 
\begin{equation}
\Gamma_{NP}(B\rightarrow K^{*}l^{+}l^{-})\,=\,
\frac{1}{2} \left(\frac{G_{F}^{2}m_{B}^{5}}{192\pi^{3}} \right)
\left(\frac{\alpha}{4\pi s_{W}^{2}} \right)^{2}
(g_{V}^{'2}+g_{A}^{'2})I_{VA}, \label{npk*}
\end{equation}
where $I_{VA}$ is the integral over the dilepton invariant mass 
($z=q^{2}/m_{B}^{2}$).  
Under the assumption that $A_{1}\approx A_{2}$, $I_{VA}$ is given by
\begin{equation}
I_{VA}=g_{V}^{2}V^{2}\int_{z_{min}}^{z_{max}}dz\frac{z}{1-z}C_{1}(z)\,\,+
\,\, g_{A}^{2}A_{1}^{2}\int_{z_{min}}^{z_{max}}dz\frac{z}{1-z}C_{2}(z),
\end{equation}
where,
\begin{eqnarray}
C_{1}(z) & = & 2\left(1+\frac{m_{K^{*}}}{m_{B}}\right)^{-2}\Phi(z) 
\nonumber \\
C_{2}(z) & = & \left[ 3 \left(1-\frac{m_{K^{*}}}{m_{B}} \right)^{2}+
\left(\frac{m_{B}}{2m_{K^{*}}}\right)^{2} 
\left(1+\frac{m_{K^{*}}}{m_{B}}\right)^{-2}
\left(z-\frac{5m_{K^{*}}^{2}}{m_{B}^{2}}\right)\Phi(z) \right]. \nonumber 
\end{eqnarray}
with 
$\Phi(z)=(1-z)^{2}+4z\left(m_{K^*}/m_B\right)^{2}$.
The limits of integration for $z$ are given by 
$z_{min}=(2m_l/m_B)^2$ and $z_{max}=(1-m_{K^{*}}/m_B)^2$.
From equation~(\ref{npk*}) we see that, the value of
$(g_{A}g_{A}^{'})^{2}$ can be determined from the measured rate of
$\Gamma(B \rightarrow K^{*} l^+ l^-)$, provided the 
value of $g_{V}^{2}(g_{V}^{'2}+g_{A}^{'2})$ is known.
For this we consider the decay of $B\rightarrow Kl^{+}l^{-}$.

The matrix element neccessary in this case is \cite{deshpande-88}
\begin{equation}
\langle K(p_{K})\left|\overline{s}\gamma_{\mu}b\right|B(p_{B})\rangle 
=(p_{B}+p_{K})_{\mu}f_{KB}^{+}(q^{2}),
\end{equation}
where again a term proportional to $q_\mu$ is dropped.
The $q^{2}$ dependence of the formfactor, again, is approximated 
by a single pole with mass $\approx m_{B}$,
\begin{equation}
f^{+}(q^{2})=\frac{f^{+}(0)}{1-q^{2}/m_{B}^2}.
\end{equation}
The decay rate is given by
\begin{equation}
\Gamma_{NP}(B\rightarrow Kl^{+}l^{-})=
g_{V}^{2}(g_{V}^{'2}+g_{A}^{'2})
\left(\frac{G_{F}^{2}m_{B}^{5}}{192\pi^{3}}\right)
\left(\frac{\alpha}{4\pi s_{W}^{2}}\right)^{2}
\left(\frac{f^{+}(0)}{2}\right)^{2}. \label{npk}
\end{equation}
We demand that the maximum value of this decay rate is the measured 
experimental value, ({\it i.e.})  
\begin{equation}
\Gamma_{exp}\, = \,\Gamma_{NP}. \label{drnp}
\end{equation}
With this assumption we calculate the upper bound on the decay rate
of $B_{s}\rightarrow l^{+}l^{-}$, arising due to $L_{VA}$, given in
Eq.~(\ref{effL}).
Using Eqs.~(\ref{npk*}), (\ref{npk}) and (\ref{drnp}), we get
\begin{equation}
g_{V}^{2}(g_{V}^{'2}+g_{A}^{'2}) = 
\frac{B_{Exp}(B\rightarrow Kl^{+}l^{-})}{
3.45\left[f^{+}(0)\right]^{2}}\times10^{4}
\end{equation}
and 
\begin{equation}
g_{A}^{2}(g_{V}^{'2}+g_{A}^{'2})=
\frac{B_{Exp}(B\rightarrow K^{*}l^{+}l^{-})\times10^{4}\,-\,
1.58V^{2}g_{V}^{2}(g_{V}^{'2}+g_{A}^{'2})}{8.94A_{1}^{2}}.
\end{equation}

In our calculation, we take the formfactors to be \cite{ali-00}
\begin{eqnarray}
f^{+}(0) & = & 0.319_{-0.041}^{+0.052} \nonumber \\
V & = & 0.457_{-0.058}^{+0.091} \nonumber \\
A_{1} & = & 0.337_{-0.043}^{+0.048},
\end{eqnarray}
and use experimental values of $B \rightarrow (K,K^*) l^+ l^-$ 
given in \cite{belle-03}.
Adding all errors in quadrature, we get
\begin{eqnarray}
g_{V}^{2}(g_{V}^{'2}+g_{A}^{'2}) & = & (1.36_{-0.44}^{+0.53})\times10^{-2}
\nonumber \\
g_{A}^{2}(g_{V}^{'2}+g_{A}^{'2}) & = & (6.76_{-3.48}^{+4.04})\times10^{-3}.
\end{eqnarray}
Thus the maximum value $(g_{A}g_{A}^{'})^{2}$ can have, is
\begin{equation}
(g_{A}g_{A}^{'})^{2}=(6.76_{-3.48}^{+4.04})\times10^{-3}
\label{maxgva}
\end{equation}

The branching ratio for $B_{s}\rightarrow l^{+}l^{-}$, 
due to $L_{VA}$, to be
\begin{eqnarray}
B(B_{s}\rightarrow e^{+}e^{-}) & = & 
1.06\times10^{-10}\cdot f_{B_{s}}^{2}(g_{A}g_{A}^{'})^{2} \nonumber \\
B(B_{s}\rightarrow\mu^{+}\mu^{-}) & = & 
4.54\times10^{-6}\cdot f_{B_{s}}^{2}(g_{A}g_{A}^{'})^{2}.
\end{eqnarray}
Substituting $f_{B_{s}} = 240\pm30$ MeV \cite{lattice} and the 
maxmimum value for $(g_{A}g_{A}^{'})^{2}$ from Eq.~(\ref{maxgva}),
we get
\begin{eqnarray}
B(B_{s}\rightarrow e^{+}e^{-}) & = & 4.06_{-2.34}^{+2.65}\times10^{-14} 
\nonumber \\
B(B_{s}\rightarrow\mu^{+}\mu^{-}) & = & 1.74_{-1.00}^{+1.13}\times10^{-9}
\end{eqnarray}
Therefore the upper bounds on the branching ratios are,
\begin{eqnarray}
B(B_{s}\rightarrow e^{+}e^{-}) & < & 6.71\times10^{-14} \nonumber \\
B(B_{s}\rightarrow\mu^{+}\mu^{-}) & < & 2.87\times10^{-9}
\end{eqnarray}
at $1 \sigma$ and 
\begin{eqnarray}
B(B_{s}\rightarrow e^{+}e^{-}) & < & 1.20\times10^{-13} \nonumber \\
B(B_{s}\rightarrow\mu^{+}\mu^{-}) & < & 5.13\times10^{-9}
\end{eqnarray}
at $3 \sigma$.

These rates are close to the SM predictions. The reason for this
is quite simple. The decay rate for an exclusive semi-leptonic process
can be written as
\begin{equation}
\Gamma = (c.c.)^2 (f.f.)^2 {\rm phase~space},
\end{equation}
where $c.c.$ is the coupling constant and $f.f.$ is the form factor.
The measured rates for the exlcusive semi-leptonic deays are close 
to the SM predictions. And we assumed that the new physics predictions 
for these processes are equal to their corresponding experimental values. 
Also, the same set of form factors are used in both SM and new physics
calculations. Thus the assumption that new physics predictions for
semi-leptonic branching ratios are equal to their experimental values
(which in turn are equal to their SM predictions) 
implies that the couplings of new physics are very close to the couplings
of the SM. This is why our new physics prediction for the purely leptonic
mode is also close to the SM prediction. Therefore, new physics,
whose effective Lagrangian for $b \rightarrow s l^+ l^-$ consists of
only vector and axial vector currents, cannot boost up the rate of
$B_s \rightarrow l^+ l^-$ due to the present experimental constraints
coming from the decays $B\rightarrow Kl^{+}l^{-}$ and 
$B\rightarrow K^{*}l^{+}l^{-}$.

For the reasons explained above, using a different set of form factors,
as for example those given in \cite{stech-00}, will not change the 
upper bound on $B_s \rightarrow l^+ l^-$ significantly. In fact, we 
find that the change is less than $10 \%$.

We can obtain a more stringent upper bound on 
$(g_{A}g_{A}^{'})^{2}$ by the following procedure.
We equate the new physics contribution for $\Gamma(B \rightarrow
(K,K^*) l^+ l^-)$ to the difference between the experimental value
and the SM contribution. This, in turn, leads to a much
more stringent upper bound on contribution of $L_{VA}$ to 
$B_s \rightarrow l^+ l^-$. In fact, at $1 \sigma$, this bound is 
consistent with $0$. At $3 \sigma$ we get 
\begin{eqnarray}
B(B_{s}\rightarrow e^{+}e^{-}) & < & 7.89\times10^{-14} \nonumber \\
B(B_{s}\rightarrow\mu^{+}\mu^{-}) & < & 3.37\times10^{-9},
\end{eqnarray}
which are again comparable to the SM predictions. 
Comparing these results with the ones obtained by previous assumption,
we see that there is not much difference in the branching ratios.
This occurs due to the relatively large errors in both the experimental
measurements and SM predictions for $\Gamma(B \rightarrow (K,K^*) l^+ l^-)$.
Thus we conclude that the presently measured values of $B \rightarrow
(K,K^*) l^+ l^-$ do not allow any large boost in the contribution of 
$L_{VA}$ to $B_s \rightarrow l^+ l^-$.

We now consider the new physics effective Lagrangian to consist of  
scalar/pseudoscalar couplings,
\begin{equation}
L_{SP}\,(b\rightarrow sl^{+}l^{-})=\frac{G_{F}}{\sqrt{2}}
\left(\frac{\alpha}{4\pi s_{W}^{2}}\right)
\bar{s}(g_{S}+g_{P}\gamma_{5})b \,
\bar{l}(g_{S}^{'}+g_{P}^{'}\gamma_{5})l. 
\label{effLs}
\end{equation}
The matrix element for the decay $B_{s}\rightarrow l^{+}l^{-}$ is given by,
\begin{equation}
M\,(B_{s}\rightarrow l^{+}l^{-})=\frac{G_{F}}{\sqrt{2}}\left(
\frac{\alpha}{4\pi s_{W}^{2}}\right)g_{P}
\langle 0\left|\overline{s}\gamma_{5}b\right|B_{s}\rangle
\left[g_{S}^{'}\bar{u}(p_{l})v(p_{\bar{l}})+
g_{P}^{'}\bar{u}(p_{l})\gamma_{5}v(p_{\bar{l}})\right]
\end{equation}
On substituting,
\begin{equation}
\langle0\left|\overline{s}\gamma_{5}b\right|B_{s}\rangle\,=\,-i\frac{f_{B_{s}}m_{B_{s}}^{2}}{m_{b}+m_{s}},
\end{equation}
we get,
\begin{equation}
M\,(B_{s}\rightarrow l^{+}l^{-})=-ig_{P}\frac{G_{F}}{\sqrt{2}}
\left(\frac{\alpha}{4\pi s_{W}^{2}}\right)
\frac{f_{B_{s}}m_{B_{s}}^{2}}{m_{b}+m_{s}}
\left[g_{S}^{'}\bar{u}(p_{l})v(p_{\bar{l}})+
g_{P}^{'}\bar{u}(p_{l})\gamma_{5}v(p_{\bar{l}})\right],
\end{equation}
where $m_{b}$ and $m_{s}$ are the masses of bottom and strange quark
respectively. Here we take the quark masses from Particle Data Group 
obtained under $\overline{MS}$ scheme \cite{pdg2004}.  
We see that in this case there is no helicity supression
$i.e.$ the rates for the decays $B_{s}\rightarrow e^{+}e^{-}$ and
$B_{s}\rightarrow\mu^{+}\mu^{-}$ will be the same provided 
$g_{P}^{'}$ and $g_{S}^{'}$ for both electrons and muons are the same. 
The calculation of the decay rate gives,
\begin{equation}
\Gamma_{NP}(B_{s}\rightarrow l^{+}l^{-})\,=\,
g_{P}^{2}(g_{S}^{'2}+g_{P}^{'2})\frac{G_{F}^{2}}{16\pi}
\left(\frac{\alpha}{4\pi s_{W}^{2}}\right)^{2}
\frac{f_{B_{s}}^{2}m_{B_{s}}^{5}}{(m_{b}+m_{s})^{2}}.
\end{equation}
The Branching ratio is given by,
\begin{equation}
B(B_{s}\rightarrow l^{+}l^{-})=
0.17\frac{f_{B_{s}}^{2}g_{P}^{2}(g_{S}^{'2}+
g_{P}^{'2})}{(m_{b}+m_{s})^{2}}\label{brbs}.
\end{equation}
To estimate the value of $g_{P}^{2}(g_{S}^{'2}+g_{P}^{'2})$, 
we again consider the related decay $B\rightarrow K^{*}l^{+}l^{-}$. 
Its matrix element, due to $L_{SP}$ is given by,
\begin{equation}
M(B\rightarrow K^{*}l^{+}l^{-})=\frac{G_{F}}{\sqrt{2}}
\left(\frac{\alpha}{4\pi s_{W}^{2}}\right)
g_{P}\langle K^{*}\left|\overline{s}\gamma_{5}b\right|B\rangle
\left[g_{S}^{'}\bar{u}(p_{l})v(p_{\bar{l}})+
g_{P}^{'}\bar{u}(p_{l})\gamma_{5}v(p_{\bar{l}})\right]
\end{equation}
as $\langle K^{*}\left|\overline{s}b\right|B\rangle=0$. The pseudoscalar
hadronic matrix element is given by \cite{chen-01}, 
\begin{equation}
\langle K^{*}\left|\overline{s}\gamma_{5}b\right|B\rangle=
-i\left(\frac{2m_{K^{*}}}{m_{b}-m_{s}}\right)A_{0}(q^{2})(q\cdot\epsilon)
\end{equation}
The $q^{2}$ dependence of the formfactor is described by a pole fit,
\begin{equation}
A_{0}(q^{2})=\frac{A_{0}(0)}{(1-q^{2}/m_{B}^{2})}.
\end{equation}
The full calculation gives,
 \begin{equation}
\Gamma_{NP}(B\rightarrow K^{*}l^{+}l^{-})=
\left(\frac{G_{F}^{2}m_{B}^{5}}{256\pi^{3}}\right)
\left(\frac{\alpha}{4\pi s_{W}^{2}}\right)^{2}
\left(\frac{2m_{K^{*}}}{m_{b}-m_{s}}\right)^{2}
\left[A_{0}(0)\right]^{2}g_{P}^{2}(g_{S}^{'2}+
g_{P}^{'2})\left(\frac{m_B}{2 m_{K^*}}\right)^{2}I_{SP}\label{npdrk*s}
\end{equation}
where,
\begin{equation}
I_{SP}=\int_{z_{min}}^{z_{max}}dz
\left[\frac{z}{\left(1-z\right)^{2}}\right]
\left[\left(1+\frac{m_{K^*}^2}{m_B^2}-z\right)^{2}-\frac{4 m_{k^*}^2}{m_B^2}
\right]^{\frac{3}{2}}.
\end{equation}
The limits of integration
for the dilepton invariant mass ($z=q^{2}/m_{B}^{2}$) are 
once again given by
$z_{min}=(2m_{l}/m_{B})^{2}$ and $z_{max}=(1-m_{K^*}/m_B)^{2}$.
Now we assume that the maximum value of this decay rate is the measured
experimental value. Thus from Eq.~(\ref{npdrk*s}), we get
\begin{equation}
g_{P}^{2}(g_{S}^{'2}+g_{P}^{'2})=
\frac{\left(m_{b}-m_{s}\right)^{2}
B_{Exp}(B\rightarrow K^{*}l^{+}l^{-})}{2.16\left[A_{0}(0)\right]^{2}}
\times10^{3}.
\end{equation}
Taking the value of $A_{0}(0)$ to be $0.471_{-0.059}^{+0.127}$ \cite{ali-00},
we get
\begin{equation}
g_{P}^{2}(g_{S}^{'2}+g_{P}^{'2})=4.02_{-1.41}^{+2.41}\times10^{-2}
\end{equation}
Substituting the value of $g_{P}^{2}(g_{S}^{'2}+g_{P}^{'2})$ 
in Eq.~(\ref{brbs}) we get,
\begin{equation}
B(B_{s}\rightarrow l^{+}l^{-})=2.10_{-0.93}^{+1.38}\times10^{-5}.
\end{equation}
The upper bound on $B(B_s \rightarrow \mu^+ \mu^-)$ from the above
equation is much higher than the present experimental upper bound
\cite{abazov-04}. Thus we see that the measured values of
$B(B \rightarrow (K,K^*) l^+ l^-)$ do not provide any useful constraint
on $L_{SP}$ contribution to $B(B_s \rightarrow \mu^+ \mu^-)$. 
The significance of this result is that if a future experiment,
such as LHC-b \cite{fortyw8}
observes $B(B \rightarrow \mu^+ \mu^-) \geq 10^{-8}$, one can
confidently assert that the {\it new physics giving rise to this large  
a branching ratio must necessarily be of scalar/psuedoscalar type}. 
Comparing the expression in 
Eq.~(\ref{brbs}) to the experimental upper bound in \cite{abazov-04}, 
we obtain the bound 
\begin{equation}
g_{P}^{2}(g_{S}^{'2}+g_{P}^{'2}) \leq 10^{-3}
\end{equation}
   
\underline{\textbf{\textit{Conclusions}}}: 
We considered the most general effective 
Lagrangian for the flavour changing neutral process 
$b \rightarrow s l^+ l^-$, arising due to new physics.
We showed that the present experimental values of 
$B(B \rightarrow (K,K^*) l^+l^-)$ set strong bounds on 
$B(B_s \rightarrow l^+ l^-)$ if the effective Lagrangian 
is product of vectors/axial-vectors. Given that the 
above semi-leptonic decay rates of B-mesons are comparable 
to their SM predicted values, we showed that 
the rate for purely leptonic decays of $B_s$ can't
be much above the their SM predicted values. We have also
derived a $3 \sigma$ upper bound on $B(B_s \rightarrow \mu^+ \mu^-) 
< 5 \times 10^{-9}$ arising from $Z'$-mediated  
flavour changing neutral currents.
If the effective Lagrangian for $b \rightarrow s l^+ l^-$ is 
product of scalars/psuedoscalars
then present experimental values of $B(B\rightarrow (K,K^*) l^+ l^-)$ 
do not lead any useful bound on $B(B_s \rightarrow l^+ l^-)$.
This leads us to the very important conclusion that, if a future
experiment observes $B_s \rightarrow l^+ l^-$ with a branching 
ratio greater than $10^{-8}$, then the new physics responsible 
for this decay must of be scalar/psuedoscalar type.

\underline{\textit {Acknowledgement:}} This work grew out of the 
discussions with Prof. Roger Forty of CERN, during WHEPP-8. We thank
Prof. Forty and other participants of WHEPP-8 for discussions on rare 
B-decays.

\end{document}